\documentclass{aa}
\usepackage{graphics}
\begin{document}

\def\Teff{$T_\mathrm{eff}$}

\thesaurus{07(04.19.1; 08.06.3; 08.23.1)}     

\title{Cool helium-rich white dwarfs from the Hamburg/ESO Survey\thanks{Based 
   on observations collected at the European 
   Southern Observatory, La Silla, Chile. This research has made use of the
   Simbad database operated at CDS, Strasbourg, France, and of the 
   Digitized Sky Survey, produced at the Space Telescope Science 
   Institute under US Government grant NAG W-2166.}
}

\author{S. Friedrich\inst1$^,$\inst2
   \and
   D. Koester\inst1
   \and
   N. Christlieb\inst3
   \and 
   D. Reimers\inst3 
   \and
   L. Wisotzki\inst3
}

\offprints{S. Friedrich}

\institute{Institut f\"ur Theoretische Physik und Astrophysik, 
  Universit\"at Kiel, 24098 Kiel, Germany
  \and
    Astrophysikalisches Institut Potsdam, An der Sternwarte 16,
  14482 Potsdam, Germany
  \and
   Hamburger Sternwarte, Universit\"at Hamburg, Gojenbergsweg 112,
       D-21029 Hamburg, Germany
}

\date{}

\maketitle

\begin{abstract}
We present an analysis of 40 cool helium-rich white dwarfs found in
the Hamburg/ESO survey. They were selected for follow-up spectroscopy 
because of their $U-B$ colour below -0.18, the absence of strong 
absorption lines, and a continuum shape similar to that of a quasar. 
Effective temperatures for individual stars were determined by 
fitting model atmospheres of nearly pure helium with a small 
admixture of hydrogen. As a consequence of the selection criteria 
all but one sample stars have \Teff\ below 20000\,K. 
Four stars clearly show helium and hydrogen lines in their spectra. 
In the spectra of another three, helium, hydrogen, and metal lines 
can be detected. For these stars hydrogen and metal abundances 
were also determined by fitting appropriate model atmospheres 
containing these elements. 
Seven sample stars most likely have helium-rich atmospheres but do not 
show any helium lines. They either have featureless spectra or show
calcium lines. 

\end{abstract}

\section{Introduction}
\begin{table}
  \caption[]{Journal of observations}
  \label{obslog}
\begin{tabular}{llllllllr}
\hline\noalign{\smallskip}
Date     & Tel.    & Instr./CCD   & Grism & Resol. \\
         &         &              & Grat. & [\AA ] \\[0.1em]
\hline\noalign{\smallskip}
Dec 1990 & 3.6\,m  & EFOSC1/RCA   & B300   & 26 \\
Apr 1991 & 3.6\,m  & EFOSC1/RCA   & B300   & 26 \\
Feb 1992 & 2.2\,m  & EFOSC2/Tek   & \#\,1  & 40 \\
Feb 1992 & 3.6\,m  & EFOSC1/Tek   & B300   & 16  \\
Apr 1992 & 3.6\,m  & EFOSC1/Tek   & B300   & 19 \\
Sep 1992 & 3.6\,m  & EFOSC1/Tek   & B300   & 21  \\
Mar 1993 & 3.6\,m  & EFOSC1/Tek   & B300   & 21  \\
Mar 1993 & 1.52\,m & B\&C/LORAL & \#\,2  & 15 \\
Feb 1994 & 1.52\,m & B\&C/LORAL & \#\,16 & 8 \\
Sep 1994 & 3.6\,m  & EFOSC1/Tek   & B300   & 21 \\
Nov 1994 & 1.52\,m & B\&C/LORAL & \#\,13 & 15  \\
Oct 1995 & 1.52\,m & B\&C/LORAL & \#\,13 & 15  \\
Mar 1996 & 1.52\,m & B\&C/\#\,39 & \#\,13 & 15 \\
Oct 1996 & 1.52\,m & B\&C/\#\,39 & \#\,13 & 15 \\
Oct 1997 & 1.52\,m & B\&C/\#\,39 & \#\,13 & 15 \\
Sep 1998 & 1.52\,m & B\&C/\#\,39 & \#\,13 & 15 \\
Nov 1998a & 1.54\,m & DFOSC/LORAL & \#\,4  & 10 \\
Nov 1998b & 1.54\,m & DFOSC/LORAL & \#\,11  & 15 \\
Nov 1998c & 1.54\,m & DFOSC/LORAL & \#\,7  & 5 \\[0.1em]
\hline
  \end{tabular}
\end{table}

\begin{table*}
  \caption[]{ Coordinates, magnitudes, cross-references and spectral types,
    if already existing, for the observed objects. Cross-references were 
    taken from the Simbad database and McCook \& Sion (\cite{mccook}). 
    Most of the EC objects are listed in Kilkenny et al. (\cite{kilkenny97}). 
    Details of individual observing campaigns can be found in Table~1. 
    The survey fields, given in column ``field'', are ESO/SERC atlas fields. 
    The $B_J$ magnitudes are accurate to better
    than 0\fm 2. \label{objects}}
    \begin{tabular}{lllllrcccc}\hline\noalign{\smallskip}
      Name & Other Names & Type & \multicolumn{2}{c}{Campaign} & \multicolumn{1}{c}{$t$ [s]}
      & Field & $\alpha$(2000) & $\delta$(2000) & $B_J$\\[0.1em]
\hline\noalign{\smallskip}
HE 0004$-$1452  & PHL665             &         & 98a & Nov 1.54m & 1800 & 607 & 00 06 36.0 & $-$14 36 14 & 17.1\\
HE 0052$-$2511  & CT195               &         & 95  & Oct 1.52m &  900 & 474 & 00 54 46.7 & $-$24 55 29 & 17.9\\
HE 0104$-$3056  & $^c$                &         & 94  & Sep 3.6m  &  300 & 412 & 01 07 12.4 & $-$30 40 16 & 17.3\\
HE 0108$-$3036  &                     &         & 92  & Sep 3.6m  &  300 & 412 & 01 11 06.3 & $-$30 20 34 & 17.7\\
HE 0122$-$2244  & KUV01223-2245       &         & 97  & Oct 1.52m &  300 & 476 & 01 24 44.6 & $-$22 29 07 & 16.8\\
HE 0127$-$3110  & BPM47178, GD1363    & DAH$^a$ & 98a & Nov 1.54m &  900 & 413 & 01 29 56.2 & $-$30 55 08 & 16.0\\
                &                     &         & 92  & Sep 3.6m  &  300 &     &            &             & \\
                &                     &         & 98c & Nov 1.54m & 1800 &     &            &             & \\
HE 0149$-$2518  & $^b$                &         & 97  & Oct 1.52m &  600 & 477 & 01 51 59.5 & $-$25 03 15 & 17.1\\
HE 0200$-$5556  &                     &         & 96  & Oct 1.52m &  300 & 153 & 02 01 57.9 & $-$55 42 16 & 17.5\\
HE 0203$-$5013  & JL281$^d$           &         & 96  & Oct 1.52m &  300 & 197 & 02 05 03.1 & $-$49 59 03 & 17.0\\
HE 0308$-$1313  & $^c$                &         & 98b & Nov 1.54m &  300 & 616 & 03 11 07.2 & $-$13 01 53 & 17.1\\
HE 0309$-$2105  & EC03097$-$2105        &         & 98c & Nov 1.54m & 1000 & 547 & 03 11 58.1 & $-$20 54 09 & 15.6\\
HE 0319$-$4344  &                     &         & 95  & Oct 1.52m &  300 & 248 & 03 21 10.1 & $-$43 34 00 & 17.7\\
HE 0413$-$3306  &                     &         & 92  & Feb 2.2m  &  300 & 360 & 04 15 20.6 & $-$32 59 11 & 15.9\\
HE 0423$-$5502  &                     &         & 95  & Oct 1.52m &  900 & 157 & 04 24 32.5 & $-$54 55 38 & 16.9\\
HE 0442$-$3027  &                     &         & 96  & Oct 1.52m & 1660 & 421 & 04 44 29.4 & $-$30 21 36 & 16.2\\
HE 0446$-$2531  &                     &         & 97  & Oct 1.52m &  480 & 485 & 04 49 01.4 & $-$25 26 36 & 16.9\\
HE 0449$-$2554  &                     &         & 98a & Nov 1.54m & 1200 & 485 & 04 51 53.8 & $-$25 49 15 & 16.5\\
                &                     &         & 94  & Nov 1.52m & 1800 &     &            &             & \\
                &                     &         & 98a & Nov 1.54m & 1200 &     &            &             & \\
HE 0453$-$4423  &                     &         & 95  & Oct 1.52m &  900 & 251 & 04 55 22.8 & $-$44 18 19 & 16.8\\
HE 0956$-$1121  & EC09565$-$1121        & DB      & 93  & Mar 3.6m  &  300 & 709 & 09 59 01.3 & $-$11 35 24 & 16.9\\
HE 1002$-$1929  &                     &         & 93  & Mar 3.6m  &  300 & 567 & 10 05 02.8 & $-$19 44 29 & 17.1\\
HE 1102$-$1850  &                     &         & 93  & Mar 3.6m  &  300 & 570 & 11 05 09.2 & $-$19 07 05 & 17.1\\
HE 1109$-$1953  & EC11091$-$1953        & DB      & 93  & Mar 3.6m  &  300 & 570 & 11 11 40.2 & $-$20 09 37 & 16.0\\
HE 1128$-$1421  & EC11285$-$1421        & DB      & 96  & Mar 1.52m &  300 & 641 & 11 31 05.5 & $-$14 38 09 & 16.7\\
HE 1128$-$1524  &                     &         & 96  & Mar 1.52m &  300 & 641 & 11 30 51.7 & $-$15 41 26 & 17.4\\
HE 1130$-$0111  &                     &         & 96  & Mar 1.52m &  300 & 858 & 11 33 25.9 & $-$01 28 30 & 17.6\\
HE 1145+0145    &                     &         & 96  & Mar 1.52m &  300 & 858 & 11 48 33.6 &   +01 28 59 & 17.2\\
HE 1146$-$0109  &                     &         & 96  & Mar 1.52m &  300 & 858 & 11 48 51.7 & $-$01 26 12 & 17.3\\
HE 1149$-$1320  & PG1149-133, EC11492 & DB2     & 92  & Feb 2.2m  &  300 & 642 & 11 51 50.5 & $-$13 37 14 & 16.1\\
HE 1214$-$2017  &                     &         & 96  & Mar 1.52m &  300 & 573 & 12 16 49.9 & $-$20 33 56 & 17.6\\
HE 1243$-$2134  & EC12436$-$2134        &         & 96  & Mar 1.52m &  300 & 574 & 12 46 18.5 & $-$21 50 57 & 16.3\\
HE 1308$-$1458  &                     &         & 91  & Apr 3.6m  &  300 & 646 & 13 11 12.9 & $-$15 14 28 & 17.2\\
HE 1350$-$1612  &                     &         & 92  & Feb 3.6m  &  300 & 649 & 13 53 34.9 & $-$16 27 05 & 17.3\\
HE 1352+0026    & PG1352+004          & DB4     & 96  & Mar 1.52m &  300 & 865 & 13 55 32.4 &   +00 11 24 & 16.1\\
HE 1409$-$1821  &                     &         & 93  & Mar 3.6m  &  300 & 578 & 14 11 48.6 & $-$18 35 04 & 15.6\\
HE 1420$-$1914  &                     &         & 96  & Mar 1.52m &  300 & 579 & 14 23 24.1 & $-$19 27 54 & 17.3\\
HE 1421$-$0856  &                     &         & 93  & Mar 3.6m  &  300 & 722 & 14 24 37.9 & $-$09 09 57 & 17.3\\
HE 1428$-$1235  & EC14289-1235  &         & 92  & Apr 3.6m  &  300 & 650 & 14 31 39.6 & $-$12 48 56 & 15.9\\
HE 2147$-$0950  & $^c$                &         & 98  & Sep 1.52m &  300 & 744 & 21 49 55.7 & $-$09 36 31 & 16.6\\
HE 2237$-$0509  & $^c$                &         & 98  & Sep 1.52m &  300 & 819 & 22 39 40.7 & $-$04 54 17 & 14.0\\
HE 2237$-$3630  & LHS 3841, LP984-76  &         & 98  & Sep 1.52m &  480 & 406 & 22 40 05.2 & $-$36 15 20 & 17.6\\
[0.1em]
\hline\noalign{\smallskip}
\multicolumn{10}{l}{$^a$ classified as magnetic WD without detectable polarization in
McCook \& Sion (\cite{mccook}), Magnetic white dwarf with a}\\
\multicolumn{10}{l}{\hspace{0.9ex} polar field strength of 176 MG 
(Reimers et al.\cite{reimersb}).}\\
\multicolumn{10}{l}{$^b$ mistakenly identified as PHL1201 in Lamontagne 
et al. \cite{lamontagne00}}\\
\multicolumn{10}{l}{$^c$ coincidences with PHL objects (PHL3345, PHL1477,
PHL166, PHL363) have been rejected using} \\ 
\multicolumn{10}{l}{\hspace{0.9ex} finding charts provided by E.~Chavira}\\
\multicolumn{10}{l}{$^d$ finding chart from Jaidee \& Lyng\aa\ 
\cite{kilkenny97}}\\
  \end{tabular}
\end{table*}

The Hamburg/ESO Survey for bright quasars (HES, Wi\-sotzki et al.
\cite{wisotzki}) was initiated in 1990 to perform a spectroscopic survey,
based on digitized objective-prism photo\-graphs taken with the ESO Schmidt
telescope, in the southern hemisphere. A nominal area of $\sim 9000$ deg$^2$ 
in the southern sky is covered with an average limiting magnitude on the 
prism plates of $B\sim 17.5$. It was conceived as a complement 
to the Hamburg Quasar Survey (HQS, Hagen et al.  \cite{hagen}), which 
aims at a complete coverage of the northern sky except for the galactic 
plane. Similar surveys have been conducted in the past to search for 
faint blue objects. Still in progress in the southern hemisphere 
are the Edinburgh-Cape survey (Stobie et al. \cite{stobie97}), the 
Homogenous Bright Quasar Survey (Cristiani et al. 
\cite{cristiani95}), and the Montreal-Cambridge-Tololo survey 
(Lamontagne et al. \cite{lamontagne00}). 

The main goal of both the HES and the HQS survey is the compilation of 
a large sample of bright QSOs. In a first step QSO candidates are 
selected by various broadband color criteria and feature detection
algorithms applied to spectra derived from low resolution scans of the prism
plates (Wisotzki et al. \cite{wisotzki00}). Since among UV 
excess objects (i.e. $U-B < -0.4$) less than 10\%\ can
be expected to be QSOs, the Hamburg Quasar Survey and the Hamburg/ESO Survey
are also a rich source of white dwarfs of various types, and early-type
subdwarfs. In a recent paper, Homeier et\clearpage al. (\cite{homeier}) present an
analysis of 80 DA white dwarfs detected through follow-up observations of the
HQS. Reimers et al. (\cite{reimersa}, \cite{reimersb}, \cite{reimersc})
reported on the detection of several new magnetic white dwarfs in the HES. A
summary of helium-rich subluminous stars can be found in Heber et al.
(\cite{heber96}) and an analysis of 47 sdO stars from the HQS in Lemke et al.
(\cite{lemke98}).

In this paper we present the analysis of 40 helium-rich white dwarfs using LTE
model atmospheres. The objects have been found in the HES and have entered the
quasar candidate sample because of $U-B\le -0.18$. Due to
the high spectral resolution of the HES spectra ($\sim 15$\,{\AA} at
H$\gamma$), it is possible to identify most of the hot stars in the quasar
candidate sample (see Reimers \& Wisotzki \cite{reimers97} for example
spectra) by their hydrogen and/or helium absorption lines. Featureless, {\em
  very\/} hot stars can be identified by their continuum shape. Therefore, the
sample of stars presented in this paper is restricted to a temperature regime
of 10000\,K$<$ \Teff\ $\le$ 20000\,K -- as the analysis shows -- in which
neither strong absorption lines of helium nor a continuum shape notably
different from that of a quasar are present. Because it was not possible to
make a definitive distinction between quasars and stars on the basis of 
the HES objective prism spectra alone, spectroscopic follow-up 
observations have been performed.

\section{Observational data}

In Table~\ref{obslog} and Table~\ref{objects} we summarize the follow-up
spectroscopy and give coordinates, magnitudes, names from other surveys, and
spectral types, if already existing, for the observed helium-rich white
dwarfs. The instruments used for spectroscopy were the ESO Faint 
Object Spectrograph and Camera (EFOSC), the Danish Faint Object 
Spectrograph and Camera (DFOSC), and the Boller \& Chivens Spectrograph 
(B\&C). A \# in Table~\ref{obslog} refers to ESO numbering of grisms, for which 
details can be found in the according telescope manuals. In general a 
spectral range between about 3800\AA\ and 7000\AA\ up to 9000\AA\ is covered 
with resolutions between about 100 to 500 \AA /mm.
The coordinates have been derived from the Digitized Sky Survey I
(DSS-I). Therefore, finding charts can easily be obtained from the online
DSS-I.  Magnitudes are given as $B_{{J}}$ magnitude, which is defined by the
sensitivity curve of the hyper-sensitized Kodak IIIa-J emulsion combined with
the filter function of a Schott BG395 filter. $B_{{J}}$ is roughly equal to
$B$ for objects around $B-V=0$ and accurate to better than $\pm$~0\fm 2 for
all sample stars.
 
\begin{figure*}
\resizebox{\hsize}{!}{\includegraphics{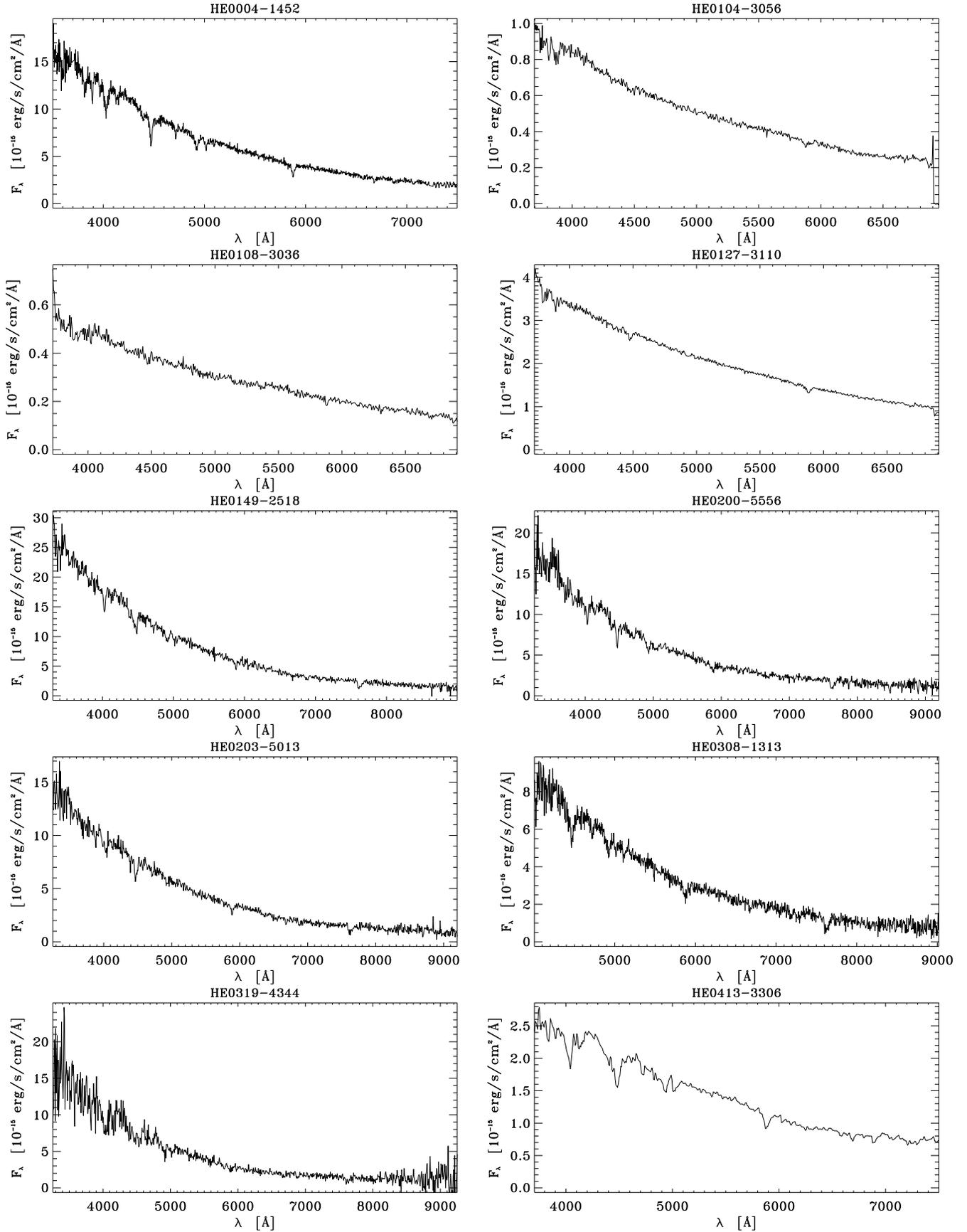}}
\caption[]{DB white dwarfs}
\label{figur1}
\end{figure*}

\begin{figure*}
\resizebox{\hsize}{!}{\includegraphics{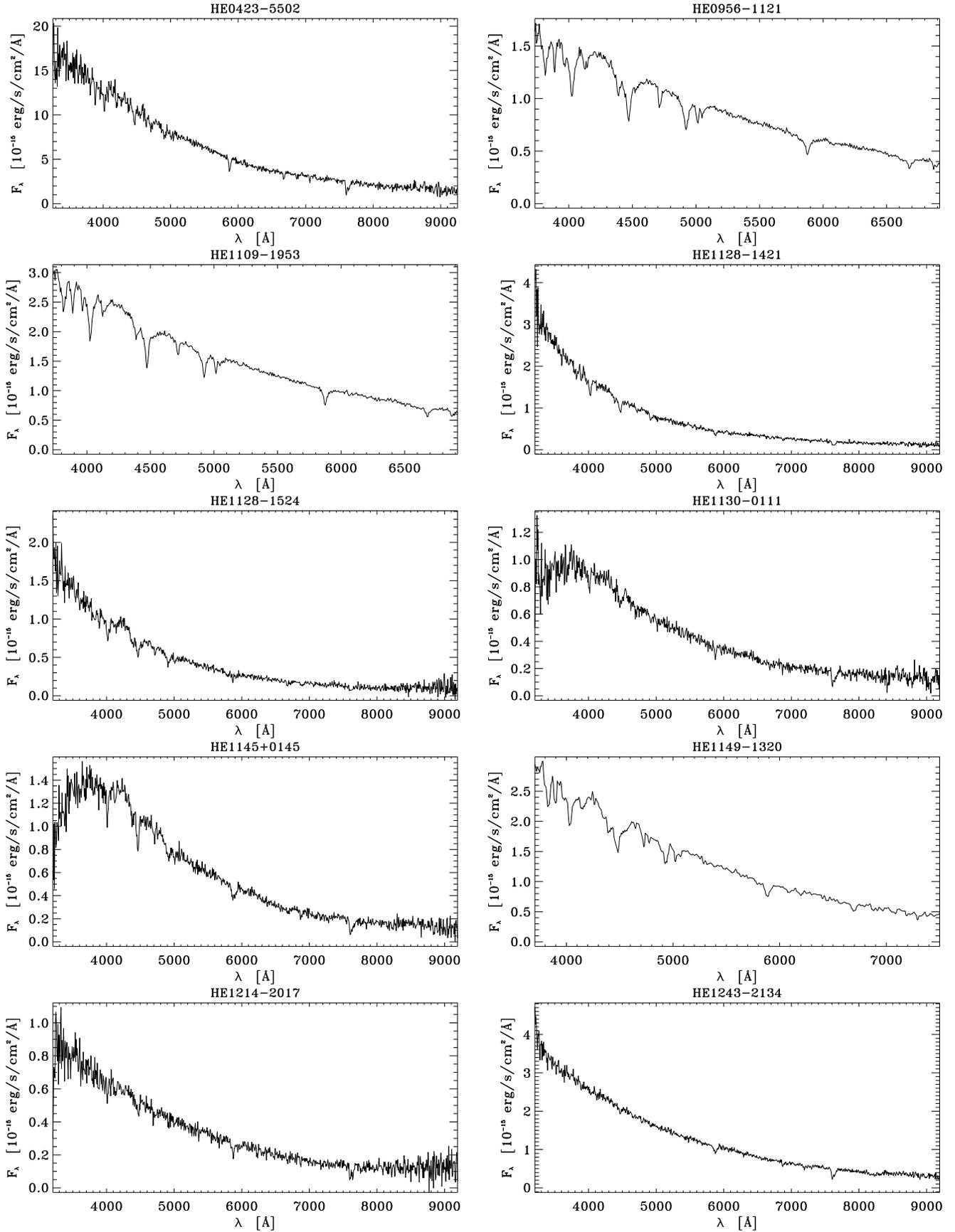}}
\caption[]{DB white dwarfs (cont.)}
\label{figur1b}
\end{figure*}

\begin{figure*}
\resizebox{\hsize}{!}{\includegraphics{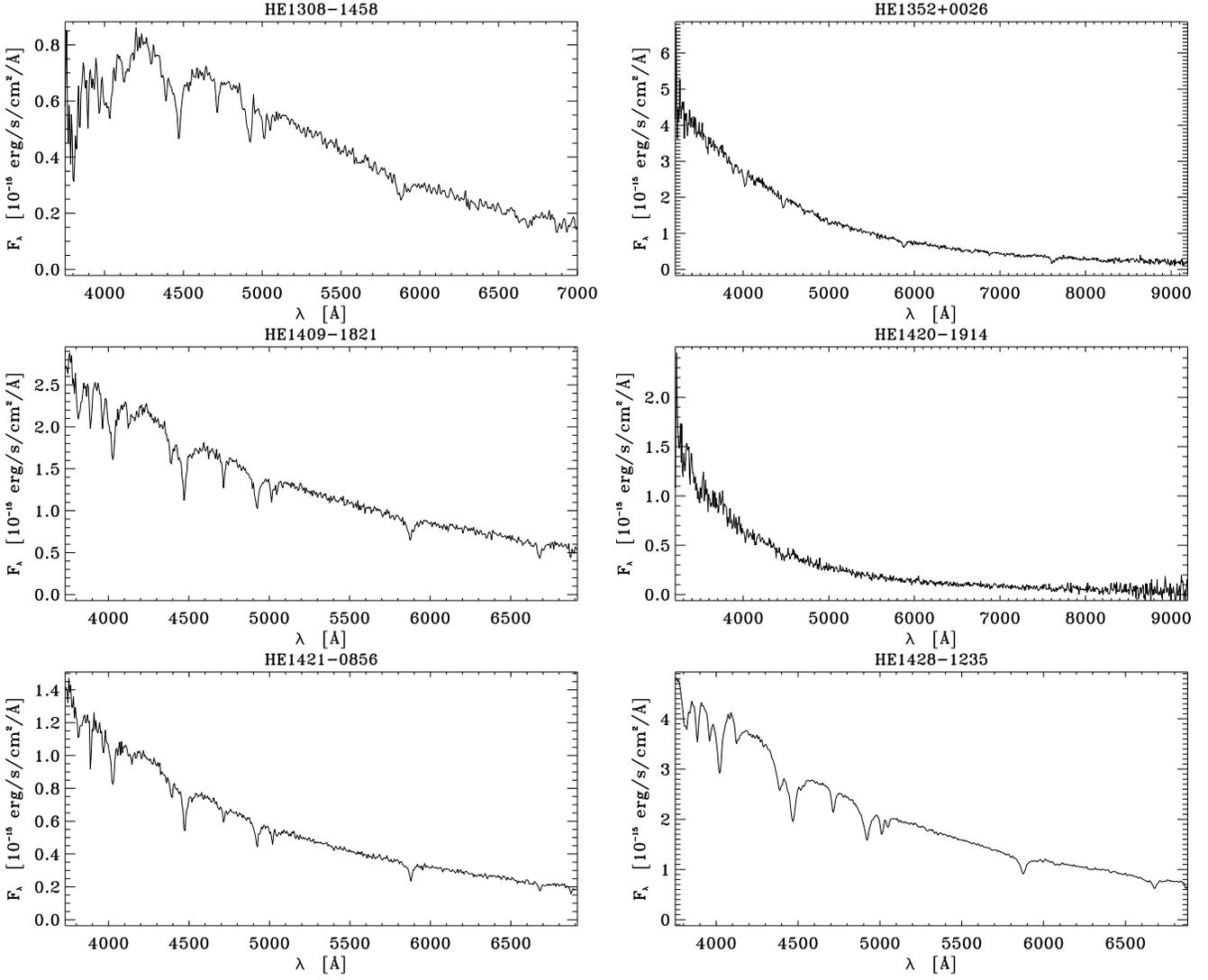}}
\caption[]{DB white dwarfs (cont.)}
\label{figur1c}
\end{figure*}

\begin{figure*}
\resizebox{\hsize}{!}{\includegraphics{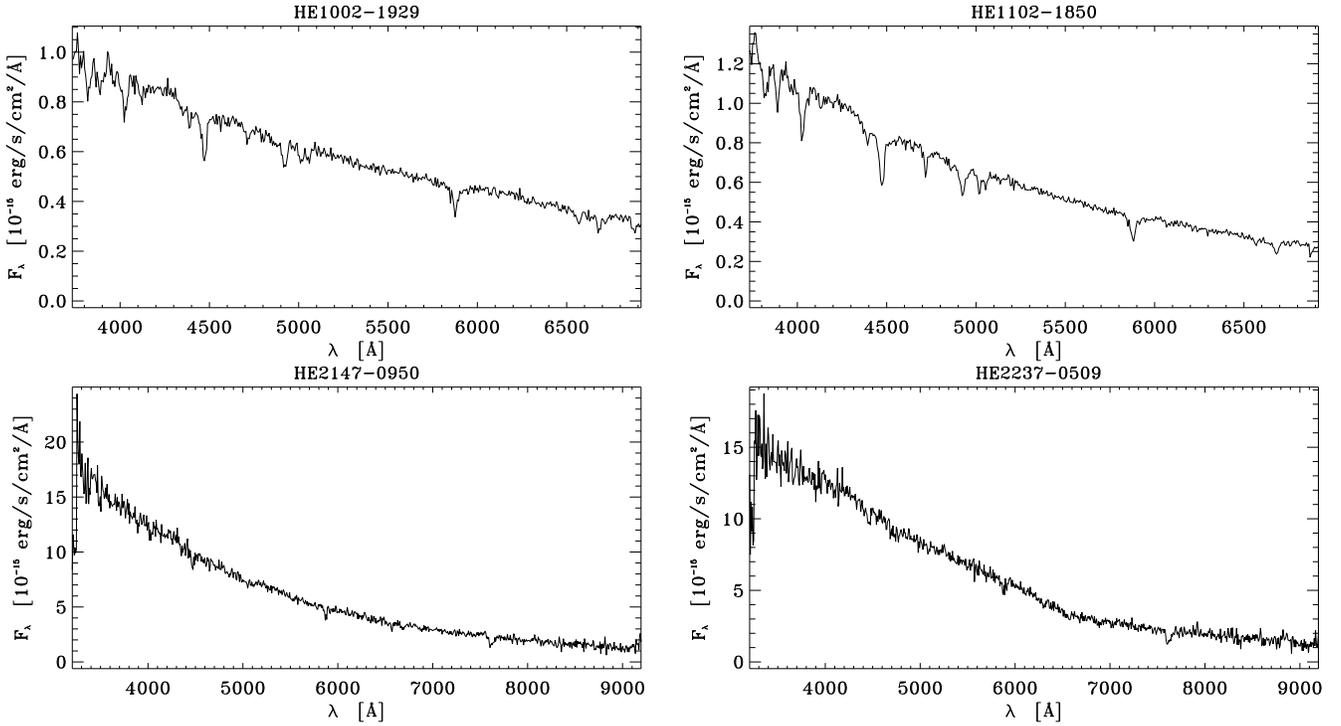}}
\caption[]{DBA white dwarfs}
\label{figur2}
\end{figure*}

\begin{figure}
\resizebox{\hsize}{!}{\includegraphics{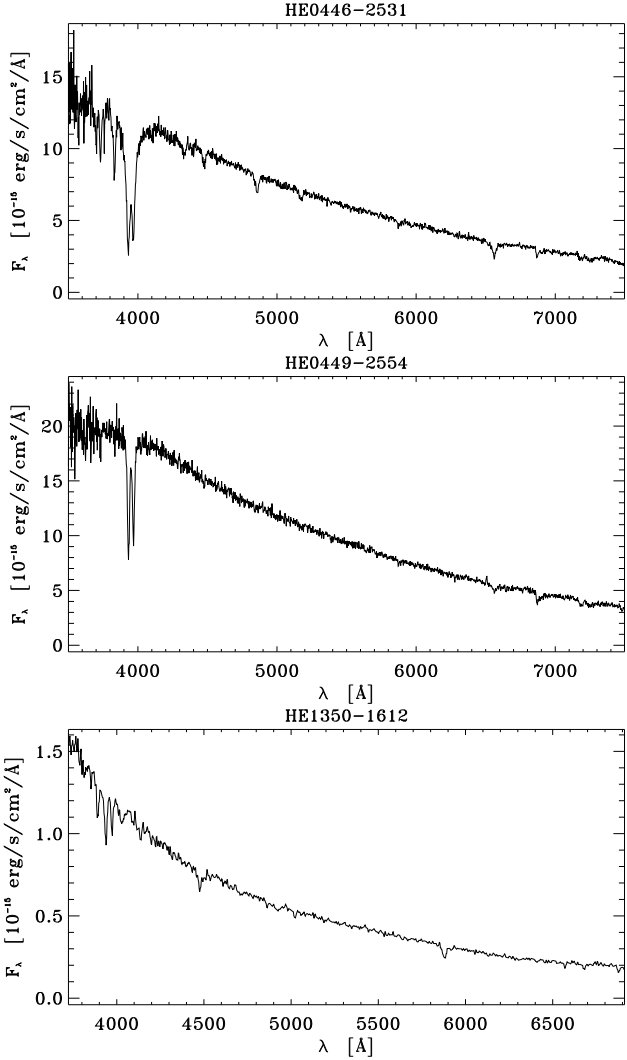}}
\caption[]{DBAZ white dwarfs}
\label{figur3}
\end{figure}

\begin{figure*}
\resizebox{\hsize}{!}{\includegraphics{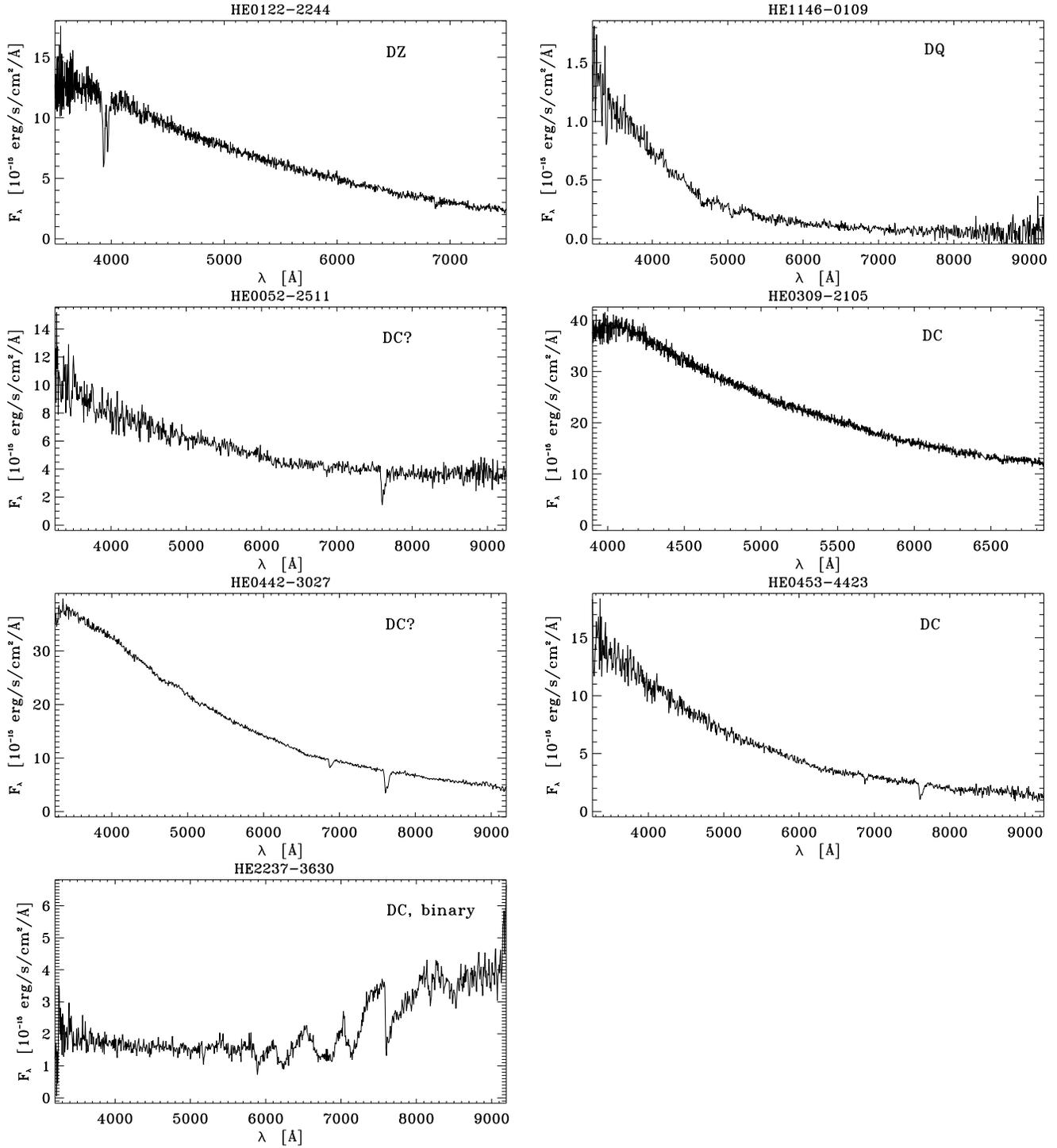}}
\caption[]{DZ, DQ, and DC white dwarfs}
\label{figur4}
\end{figure*}

The spectra of all DB white dwarfs are depicted in Fig.~\ref{figur1} to
\ref{figur1c}. The spectra of DBA and DBAZ white dwarfs can be found in 
Fig.~\ref{figur2} and Fig.~\ref{figur3}, respectively, and the spectra 
of the DZ, DQ, and DC white dwarfs are summarized in Fig.~\ref{figur4}.

\section{Spectral analysis}

\subsection{Models}

In order to derive effective temperatures the observed spectra have been
fitted to model spectra of nearly pure helium atmospheres with a small
admixture of hydrogen ($10^{-6}$ relative to helium by number) covering a
temperature range from 10000\,K to 35000\,K in 250 K steps up to 16000\,K, in
500\,K steps between 16000\,K and 20000\,K, and 1000\,K steps above 
20000\,K. $\log g$ was given between 7.75 and 8.5 in 0.25 dex increments. 
Between 16500\,K and
20000\,K we extended the grid to $\log g = 9$. \ion{He}{i} line profiles were
calculated with our LTE atmosphere codes (see Finley et al. \cite{finley} for
a recent description) in which the improved \ion{He}{i} line broadening theory
of Beauchamp et al. (\cite{beauchamp}) was implemented. These Stark profiles
have been calculated for temperatures ranging from 10000\,K to 40000\,K and
consider broadening by ions and electrons. Line broadening by neutral
perturbers, which becomes important at lower temperatures, was not included in
our calculation of line profiles. This might lead to some uncertainties.
However, a temperature determination from the continuum slopes of
HE~0446$-$2531 and HE~0449$-$2554, for which we have absolutely calibrated
flux spectra, showed the deviation in \Teff\ to be less than 1000\,K at
temperatures of about 12000\,K.

\subsection{Determination of \Teff }


Effective temperature and $\log g$ were determined in a $\chi^2$ 
fitting procedure based on a Levenberg-Marquard algorithm 
(cf. Press et al. \cite{press}) by comparison of \ion{He}{i}
line profiles with synthetic spectra from the He/H atmospheres. 
The method is very similar to that described in detail in Homeier 
et al. (\cite{homeier}). 

We started the fitting procedure with both temperature and $\log g$ as a free
parameter. However, at the given data quality the dependence on $\log g$
turned out to be rather small, and it was thus not possible to determine it
unambiguously. In most cases a higher $\log g$ could be compensated by a
higher temperature, and vice versa, with roughly the same $\chi^2$ value.
Furthermore, systematic effects like small differences in the starting values
for \Teff\ and $\log g$ might change the solution by much more than the
statistical errors.  We therefore determined \Teff\ for $\log g$ fixed to 8,
too.  Comparison of temperatures derived from fits with $\log g$ as free
parameter and fixed to 8 only revealed small differences, which could often be
regarded as equal within their statistical errors.  This is reflected by the
mean of all fitted $\log g$ values of 8.15$\pm$0.24. In Table~\ref{temp} we
have compiled the results of our analysis. The given errors for temperatures
and $\log g$ are formal statistical errors from the covariance matrix of the
fit, which do not reflect any systematic errors. As has been discussed by many
authors using similar methods (see e.g. Homeier et al. \cite{homeier};
Napiwotzki et al. \cite{napi}), external errors can be higher by a large
factor.

\subsection{Determination of hydrogen and metal abundances}

As for the DB stars, \Teff\ was determined for DBA and DBAZ white dwarfs by
comparing \ion{He}{i} line profiles with those from synthetic spectra of our
He/H atmospheres. We did not perform selfconsistent calculations with hydrogen
and metal lines already included in the atmospheres. In the next step model
atmospheres for the derived effective temperatures were calculated which
contain calcium and/or hydrogen in estimated amounts for the respective star.
The resulting temperature and pressure stratification was then used to compute
detailed synthetic spectra with varying calcium and/or hydrogen abundances.
For HE~0446$-$2531 also magnesium and iron were considered. Abundances were
then obtained by comparison of observed equivalent widths and line profiles to
those from the synthetic spectra. Unless otherwise mentioned, equivalent
widths of H$\alpha$, H$\beta$, and H$\gamma$ were determined between
6543\,{\AA} and 6583\,{\AA}, 4841\,{\AA} and 4881\,{\AA}, and 4310\,{\AA} and
4370\,{\AA}, respectively; equivalent widths of the \ion{Ca}{ii} doublet were
determined between 3890\,{\AA}\ and 3990\,{\AA}.

When comparing HE~0446-2531 and HE~0449-2554, the great difference in calcium
abundances (approximately a factor of 30) is surprising because temperatures
are similar, and equivalent widths differ only by a factor of 2. However, their
hydrogen abundances also differ by about a factor of 20. This affects the
atmospheric structure since hydrogen contributes electrons, and changes the
opacity. In turn the changed atmospheric structure does influence the line
spectrum.

\section{Discussion of individual objects}

\subsection{DB white dwarfs}

\subsubsection{HE~0127$-$3110}
HE~0127$-$3110 was observed several times during the follow-up observations 
performed for the HES: Spectra were obtained in September 1992 (used here), 
September 1994, November 1994, and November 1998. Spectra 
from 1994 show broad absorption troughs, varying in depth and profile, 
whereas the September 1992 spectrum resembles a DB spectrum. After a 
careful inspection of individual telescope pointings to exclude the 
accidental observation of a different star, we conclude that 
HE~0127$-$3110 is variable.  

The only common feature of all spectra is an absorption line at 5876\,{\AA} 
which can be attributed to \ion{He}{i}. The broad absorption features in the 
spectra of 1994 led to the classification as a magnetic DA white dwarf 
with a polar field strength of 176 MG (Reimers et al. \cite{reimersb}). 
In this scenario the putative \ion{He}{i} line at 5876\,{\AA} is due 
to a hydrogen H$\alpha$ component ($2s_0\rightarrow 3p_0$), which 
becomes stationary only at a field strength of 235 MG with a minimum 
wavelength of 5830\,{\AA}. It is interesting to mention that HE~2201$-$2250, 
whose spectrum looked almost identical to the 1994 spectra of 
HE~0127$-$3110 (Reimers et al. \cite{reimersb}), meanwhile also changed its 
spectroscopic appearance to a DB white dwarf. A detailed discussion 
will be published in a forthcoming paper.

\subsection{DBA white dwarfs}

\begin{table*}
  \caption[]{Results from the temperature and $\log g$ determination and 
             derived spectral types for the analyzed stars. Equivalent 
             widths are given in \AA ngstr\"om }
  \label{temp}
\begin{tabular}{llllll}\hline\noalign{\smallskip}
Name & Type & \Teff\ & $\log g$ & \Teff\ for & Remarks \\
     &      &        &          & $\log g=8$ &\\[0.1em]
\hline\noalign{\smallskip}
HE 0004$-$1452 &   DB & 16800$\pm$100 & 8.17$\pm$0.06 & 16700$\pm$100 &\\
HE 0104$-$3056 &   DB & 12600$\pm$320 & 8.11$\pm$0.30 & 12500$\pm$580 & \\
HE 0108$-$3036 &   DB & 13600$\pm$260 & 8.36$\pm$0.21 & 13200$\pm$430 &\\
HE 0127$-$3110 &   DB & 13800$\pm$120 & 8.48$\pm$0.09 & 13500$\pm$130 & 
variable (?), possibly magnetic\\
HE 0149$-$2518 &   DB & 16900$\pm$280 & 8.21$\pm$0.17 & 16500$\pm$310 & \\
HE 0200$-$5556 &   DB & 18500$\pm$250 & 8.26$\pm$0.12 & 17900$\pm$230 & \\
HE 0203$-$5013 &   DB & 16200$\pm$310 & 8.37$\pm$0.20 & 16000$\pm$290 & \\
HE 0308$-$1313 &   DB & 18000$\pm$170 & 8.90$\pm$0.20  &18000$\pm$350 &\\
HE 0319$-$4344 &   DB & 18400$\pm$710 & 7.82$\pm$0.39 & 18600$\pm$800 & \\
HE 0413$-$3306 &   DB & 16600$\pm$70 & 7.80$\pm$0.05 & 16700$\pm$70 & \\
HE 0423$-$5502 &   DB & 15600$\pm$200  & 8.12$\pm$0.14 & 15500$\pm$200 & \\
HE 0446$-$2531 & DBAZ & 12600$\pm$250 & 7.77$\pm$0.23 & 12600$\pm$280 & 
W($\lambda$)=$50.7^{+5}_{-3}$/$54.3\pm5$ (\ion{Ca}{ii}, Oct97/Nov98),\\
&&&&&W($\lambda$)=$8.6\pm1$ (H$\alpha$), 
W($\lambda$)=$4.7\pm 0.5$ (H$\beta$),\\
&&&&&W($\lambda$)=$2.9^{+0.5}_{-1.0}$ (H$\gamma$)\\
HE 0449$-$2554 & DBAZ & 12500$\pm$220 & 8.23$\pm$0.15 & 12200$\pm$290 & 
W($\lambda$)=$25.0^{+2}_{-3}$/$22.8^{+2}_{-3}$ (\ion{Ca}{ii}, Nov94/Nov98)\\
&&&&&W($\lambda$)=$2.8\pm 0.5$ (H$\alpha$)\\
HE 0956$-$1121 &   DB & 18800$\pm$90 & 8.02$\pm$0.04 & 18400$\pm$80 & \\
HE 1002$-$1929 &  DBA & 16500$\pm$100 & 8.26$\pm$0.06 & 16200$\pm$100 & 
W($\lambda$)=$3.6\pm1$ (H$\alpha$)\\
HE 1102$-$1850 &  DBA & 17000$\pm$100 & 8.25$\pm$0.06 & 16700$\pm$120 & 
W($\lambda$)=$2.4\pm 0.6$ (H$\alpha$), W($\lambda$)=$1.1\pm0.6$ (H$\beta$)\\
HE 1109$-$1953 &   DB & 17100$\pm$70 & 7.85$\pm$0.04 & 17000$\pm$60 & \\
HE 1128$-$1421 &   DB & 16900$\pm$240 & 8.04$\pm$0.14 & 16600$\pm$240 & \\
HE 1128$-$1524 &   DB & 26800$\pm$1420 & 8.33$\pm$0.16 & 27800$\pm$1120 & \\
HE 1130$-$0111 &   DB & 15000$\pm$300 & 8.00$\pm$0.19 & 15100$\pm$330 &\\
HE 1145+0145 &   DB & 16900$\pm$250 & 8.02$\pm$0.15 & 16600$\pm$250 & \\
HE 1149$-$1320 &   DB & 17300$\pm$120 & 8.07$\pm$0.07 & 17600$\pm$130 & \\
HE 1214$-$2017 &   DB & 16700$\pm$200 & 8.05$\pm$0.12 & 15700$\pm$210 & \\
HE 1243$-$2134 &   DB & 13800$\pm$240 & 8.44$\pm$0.15 & 13300$\pm$280 & \\
HE 1308$-$1458 &   DB & 19500$\pm$190 & 8.35$\pm$0.07 & 18800$\pm$170 & \\
HE 1350$-$1612 & DBAZ & 15000$\pm$80 & 8.28$\pm$0.06 & 14600$\pm$90 & 
W($\lambda$)=$8\pm1$ (\ion{Ca}{ii}),\\ 
&&&&& W($\lambda$)=$2.2^{+0.3}_{-0.5}$ (H$\alpha$), 
W($\lambda$)=$1\pm0.5$ (H$\beta$)\\
HE 1352+0026 &   DB & 15700$\pm$150 & 8.09$\pm$0.10 & 15400$\pm$170 & \\
HE 1409$-$1821 &   DB & 18400$\pm$110 & 7.88$\pm$0.06 & 18300$\pm$130 & \\
HE 1420$-$1914 &   DB & 17900$\pm$220 & 8.24$\pm$0.12 & 17200$\pm$370 & \\
HE 1421$-$0856 &   DB & 17100$\pm$90 & 7.84$\pm$0.05 & 17200$\pm$90 & \\
HE 1428$-$1235 &   DB & 19500$\pm$50 & 7.87$\pm$0.02 & 19500$\pm$60 & \\
HE 2147$-$0950 &  DBA & 14600$\pm$220& 8.17$\pm$0.13 & 14100$\pm$410 & 
W($\lambda$)=$3.5\pm0.5$ (H$\alpha$)\\
HE 2237$-$0509 &  DBA  & 13300$\pm$360& 8.34$\pm$0.27& 12900$\pm$380 & 
W($\lambda$)=$3.4\pm1$ (H$\alpha$)\\
[0.1em]
\hline
&&&&&\\
\multicolumn{6}{l}{Possible helium-rich stars without helium lines}\\[0.3em]
\hline\noalign{\smallskip}
HE 0052$-$2511 &      & --- & --- & --- & possibly DC or magnetic\\
HE 0122$-$2244 &   DZ & --- & --- & $10000^{+2000}_{-1000}$ & 
W($\lambda$)=$18.4^{+3}_{-5}$/$19.0\pm 2$ (\ion{Ca}{ii}, Oct97/Nov98)\\
HE 0309$-$2105 &   DC & --- & --- & --- & \\
HE 0442$-$3027 &      & --- & --- & --- & possibly DC or DQ\\
HE 0453$-$4423 &   DC & --- & --- & --- & \\
HE 1146$-$0109 &   DQ & --- & --- & --- & C$_2$ bands\\
HE 2237$-$3630 &   DC & --- & --- & --- & binary\\[0.1em]
\hline
  \end{tabular}
\end{table*}

\subsubsection{HE~1002$-$1929 and HE~1102$-$1850}

In the spectra of HE~1002$-$1929 and HE~1102$-$1850 the H$\alpha$ line is 
clearly visible with equivalent widths of $(3.6\pm 1)$\,{\AA} and 
$(2.4\pm 0.6)$\,{\AA}, respectively. The H$\beta$ line 
(W($\lambda$)=$(1.1\pm 0.6)$\,{\AA}) can be detected only in the 
spectrum of HE~1102$-$1850, despite the larger equivalent width of 
H$\alpha$ in HE~1002$-$1929 and a similar temperature. 
From the equivalent widths we calculated hydrogen abundances of 
$(5^{+5}_{-2})\cdot 10^{-5}$ and $(2^{+2}_{-1})\cdot 10^{-5}$ for 
HE~1002$-$1929 and HE~1102$-$1850, respectively. Metal lines could not be 
detected.

\subsubsection{HE~2147$-$0950 and HE~2237$-$0509}
HE~2147$-$0950 and HE~2237$-$0509 show absorption features at the position 
of H$\alpha$ with equivalent widths of $(3.5\pm0.5)$\,{\AA} and 
$(3.4\pm1.0)$\,{\AA}, respectively. Because both spectra are very noisy, 
this should be taken with caution. The measured equivalent widths would 
correspond to hydrogen abundances of $(4\pm2)\cdot 10^{-5}$ and 
$(6\pm4)\cdot 10^{-5}$, respectively.

\subsection{DBAZ white dwarfs}

\subsubsection{HE~0446$-$2531}
The most prominent lines in the November 1998 spectrum of HE~0446$-$2531 are
hydrogen lines of the Balmer series up to H$\gamma$, the \ion{He}{i} line at
4471\,{\AA} and a somewhat weaker \ion{He}{i} line at 5876\,{\AA}.  The
October 1997 spectrum is too noisy to let the Balmer lines be discerned.
Equivalent widths for H$\alpha$ and H$\beta$ were determined between
6530\,{\AA} and 6590\,{\AA}, and 4820\,{\AA} and 4885\,{\AA}, respectively.
Together with the equivalent width of H$\gamma$ they result in a hydrogen
abundance of $(7^{+3}_{-2})\cdot 10^{-4}$.  The equivalent width of the
calcium doublet was determined between 3850\,{\AA} and 4050\,{\AA} to comprise
the whole doublet. This is justified because the \ion{He}{i} line at
4026\,{\AA} is to weak to significantly contribute. The calcium abundance 
derived from an equivalent width of $(52.5^{+5}_{-6})$\,{\AA} 
(unweighted mean from the October 97 and November 98 spectrum) 
amounts to $(2\pm 1)\cdot 10^{-6}$.
At 3704\,{\AA}, 3735\,{\AA}, 3834\,{\AA}, and 5170\,{\AA} lines can be
detected, which are most probably due to neutral magnesium and neutral as well
as ionized iron. With this assumption we derive abundances of
$(2^{+3}_{-1})\cdot 10^{-6}$ and $(2\pm 1)\cdot 10^{-6}$ for magnesium and
iron, respectively.

\subsubsection{HE~0449$-$2554}
Since the November 94 spectrum is very noisy HE~0449-2554 was reobserved in 
November 1998. Besides the  doublet of calcium at 3934\,{\AA} and 3968\,{\AA} 
already seen in the November 94 spectrum, the new spectrum clearly shows  
an H$\alpha$ line. At 4471\,{\AA} and 5876\,{\AA} also weak features 
can be spotted. 
They are consistent with line depths and line strengths of corresponding 
\ion{He}{i} lines in a synthetic spectrum with \Teff\ = 12200\,K. The 
equivalent width of the H$\alpha$ line ($(2.8\pm 0.5)$\,{\AA}) corresponds 
to a hydrogen abundance of $(3\pm1)\cdot 10^{-5}$. The unweighted mean 
equivalent width of the calcium doublet from both observing runs 
($23.9^{+2}_{-3}$\,{\AA}) was taken to determine a calcium 
abundance of $(7\pm 2)\cdot 10^{-8}$.

\subsubsection{HE~1350$-$1612}
In addition to \ion{He}{i} lines also H$\alpha$, H$\beta$, and the \ion{Ca}{ii}
doublet at 3934\,{\AA} and 3968\,{\AA} can be detected in the spectrum.
With an effective temperature of 14600$\pm$90\,K, this star is
one of the very few DBAZ stars in the temperature range of about 15000\,K,
where \ion{He}{i} lines make temperature determinations reliable. The only
other such star known is HS~2253+8023 (Friedrich et al.  \cite{friedrich};
\cite{friedrichb}), and perhaps GD40, with a published equivalent width of
H$\alpha$ of 1\,{\AA} (Greenstein \& Liebert \cite{greenstein}). Hydrogen and
calcium abundances determined from equivalent widths amount to $(2\pm
1)\cdot10^{-5}$ and $(2^{+1.0}_{-1.5})\cdot 10^{-7}$, respectively.

\subsection{DZ white dwarfs}

\subsubsection{HE~0122$-$2244}

The \ion{Ca}{ii} doublet at 3934\,{\AA}/3968\,{\AA} and an otherwise
featureless spectrum classify HE~0122$-$2244 as a DZ white dwarf. 
An approximate temperature of $10000^{+2000}_{-1000}$\,K for 
$\log g = 8$ was determined from
the continuum slope of the photometric November 98 spectrum. For this
temperature the equivalent width of the \ion{Ca}{ii} doublet
($18.7^{+4}_{-5}$\,{\AA}, unweighted mean) corresponds to a calcium abundance
of $(4^{+1}_{-2})\cdot10^{-10}$.

\subsection{DC and DQ white dwarfs}

\begin{table}
\setbox\strutbox=\hbox{\vrule height9.5pt depth4.5pt width0pt}
  \caption[]{Element abundances (relative to helium) for the DBA, DBAZ, 
and DZ white dwarfs}
  \label{abunda}
\begin{tabular}{llcccc}
\hline
object & Type & H & Mg & Ca & Fe \\
       &      & $10^{-5}$ & $10^{-6}$ & $10^{-8}$ & $10^{-6}$\\[0.1em]
\hline\noalign{\smallskip}
HE 0122$-$2244 & DZ & & & $0.04^{+0.01}_{-0.02}$ & \\
HE 0446$-$2531 & DBAZ & $70^{+30}_{-20}$ & $2^{+3}_{-1}$ & $200\pm 100$ & $2\pm 1$ \\
HE 0449$-$2554 & DBAZ & $3\pm 1$ & & $7\pm 2$ & \\
HE 1002$-$1929 & DBA & $5^{+5}_{-2}$ & & & \\
HE 1102$-$1850 & DBA & $2^{+2}_{-1}$ & & & \\
HE 1350$-$1612 & DBAZ & $2\pm 1$ & & $20^{+10}_{-15}$ & \\
HE 2147$-$0950 & DBA & $4\pm 2$ & & & \\
HE 2237$-$0509 & DBA & $6\pm 4$ & & & \\[0.1em]
\hline
\end{tabular}
\end{table}

\subsubsection{HE~0052$-$2511}
Weak flux depressions at 5000\,{\AA} and 6300\,{\AA} are the only features in
the observed spectral range from 3500\,{\AA} to 9200\,{\AA}. Therefore, this
star might be of type DC, or possibly a magnetic white dwarf. If this star is
magnetic, the main constituent of its atmosphere could be helium, because
helium, contrary to hydrogen, has no stationary components and hence has 
no detectable features over a wide range of field strengths (Schmelcher, 
priv.comm.).

\subsubsection{HE~0309$-$2105}
HE~0309$-$2105 was already observed by Sefako et al. (\cite{sefako99}). 
On the basis of its colours ($B-V=0.10\pm 0.05$, $U-B=-0.08\pm 0.05$) 
derived from $UBV$ photometry they were 
not able to classify the object. However, our synthetic colours derived from 
prism plates of the HES and confirmed by spectroscopy yield 
$B-V=-0.10\pm 0.08$ and $U-B=-0.88\pm 0.08$, respectively, which place 
HE~0309$-$2105 well in the regime of white dwarfs in the two-colour diagram. 
Together with its featureless spectrum it is therefore classified as DC.  

\subsubsection{HE~0442$-$3027}
C$_2$ bands might be the reason for weak flux depressions at 
4700\,{\AA} and 5100\,{\AA}. Therefore, HE~0442$-$302 might rather be a DQ 
than a DC.

\subsubsection{HE~1146$-$0109}
HE~1146$-$0109 displays strong C$_2$ bands at 4670\,{\AA} and 5140\,{\AA}. 
Since no other features can be detected in the spectrum, the star is 
classified as DQ.

\subsubsection{HE~2237$-$3630}
This star has a composite spectrum, which shows no clear features bluewards 
of 5800\,{\AA}, and strong TiO bands redwards of 5800\,{\AA}. There are hints
for absorption features at 3730\,{\AA} and 4026\,{\AA}, which could be attributed
to helium, and one at 5170\,{\AA}, which could possibly be due to magnesium 
and iron. We assume that HE~2237$-$3630 is a binary star with a helium-rich 
white dwarf component and a M-dwarf component.

\section{Discussion}
We have presented a model atmosphere analysis of cool helium-rich white dwarfs
found in the Hamburg/ESO survey for which follow-up spectroscopy has been
performed. Since the selection criterium for these stars was confined to a
continuum slope similar to QSOs, and no strong absorption lines, this sample
of helium-rich white dwarfs is biased towards \Teff\ below 20000\,K. 

Among the 33 analyzed DB stars, we found 4 DBA, if we include HE~2147$-$0950
and HE~2237$-$0509 with a somewhat uncertain identification of H$\alpha$, and
3 DBAZ stars; hence, a fraction of 21\% in our sample shows traces of hydrogen.
This is consistent with the findings of Shipman et al. (\cite{shipman87}), who
found 19\% DBA white dwarfs in their complete sample of cool DB stars from the
Palomar-Green survey.

Especially the DBAZ stars are important in the context of the ongoing
discussion of accretion of hydrogen and metals from the interstellar medium.
Since the diffusion times of metals in helium-rich atmospheres are short
compared to evolutionary time scales (Paquette et al. \cite{paquette}), it is
generally accepted that metals cannot be of primordial origin but must be
accreted from the interstellar medium. Compared to the accretion-diffusion
model of Dupuis et al. (\cite{dupuisa}, \cite{dupuisb}, \cite{dupuisc}), which
predicts element abundances in cool white dwarf atmospheres in consideration
of accretion from the interstellar medium, the derived Ca abundances for the
DBAZ stars found and the magnesium and iron abundance for HE~0446$-$2531 are
consistent with accretion in solar element proportions at high accretion
rates. For HE~0446$-$2531 it is also possible to determine metal-to-metal
ratios. Whereas the Fe/Mg ratio falls within the predicted range, the Fe/Ca
and Mg/Ca ratios are below their respective lower limits, but given the large
errors they might still be compatible with accretion in solar element
proportions.
 
Contrary to the metals, which sink down in the atmospheres and disappear very
rapidly, hydrogen has the tendency to float on top of a helium-rich
atmosphere. It is therefore accumulated with time in the surface layers. Thus
one would expect a metal-to-hydrogen ratio, which is below the solar value. As
already observed in other helium-rich metal-line white dwarfs (see Dupuis et
al. \cite{dupuisc} for a compilation) the Ca/H ratio is larger by a factor of
$10^3$ to $10^4$ than the solar value for the new DBAZ stars, supporting the
hypothesis that accretion of hydrogen is somehow reduced compared to accretion
of metals. The mechanism most widely discussed for this ``hydrogen screening''
is the propeller mechanism of Illarionov \& Sunyaev (\cite{illarionov}),
originally developed for accretion of interstellar matter onto neutron stars.
Wesemael \& Truran (\cite{wesemael82}) first discussed it in the context of
white dwarfs. They proposed that metals are accreted in the form of grains
onto a slowly rotating, weakly magnetized ($10^5$~G) white dwarf, whereas
ionized hydrogen is repelled at the Alfv\'en radius. However, it remains to be
demonstrated that the necessary conditions for rotation, magnetic field and
hydrogen-ionizing radiation are fulfilled to make the propeller mechanism
operate.

\begin{acknowledgements}
The Hamburg/ESO Survey has been supported by the DFG under grants Re~353/33 
and Re~353/40.
Work on the Hamburg Quasar Survey in Kiel is supported by a grant 
from the DFG under Ko~738/10-3. We want to thank Dr.~E.~Chavira (INAOE)
for supplying finding charts of PHL objects. D.K. is grateful to 
Dr.~Jay Holberg (Lunar and Planetary Laboratory) and Dr.~Jim Liebert 
(Steward Observatory) for their hospitality during his sabbatical at 
the University of Arizona.
\end{acknowledgements}


\begin{thebibliography}{}
\bibitem[1997]{beauchamp}
Beauchamp, A., Wesemael, F., Bergeron, P., 1997, ApJS 108, 559
\bibitem[1995]{cristiani95}
Cristiani, S., La Franca, F., Andreani, P., et al., 1995, A\&AS 112, 347
\bibitem[1992]{dupuisa}
Dupuis, J., Fontaine, G., Pelletier, C., Wesemael, F., 1992, ApJS 82, 505
\bibitem[1993a]{dupuisb}
Dupuis, J., Fontaine, G., Pelletier, C., Wesemael, F., 1993a, ApJS 84, 73
\bibitem[1993b]{dupuisc}
Dupuis, J., Fontaine, G., Wesemael, F., 1993b, ApJS 87, 345
\bibitem[1997]{finley}
Finley, D.S., Koester, D., Basri, G., 1997, ApJ 488, 375
\bibitem[1999a]{friedrich}
Friedrich, S., Koester, D., Heber, U., Reimers, D.: 1999,
Analysis of UV and optical spectra of helium-rich white dwarfs
with trace elements. In: Proceedings of the 11th European wokshop
on white dwarfs, J.-E. Solheim \& E. Mei\v stas (eds.),
ASP Conf. Ser., vol. 169, 505
\bibitem[1999b]{friedrichb}
Friedrich, S., Koester, D., Heber, U., Jeffery, C.S., Reimers, D., 
1999, A\&A 350, 865 
\bibitem[1990]{greenstein}
Greenstein, J.L., Liebert, J.W., 1990, ApJ 360, 662
\bibitem[1995]{hagen}
Hagen, H.-J., Groote, D., Engels, D., Reimers, D., 1995, A\&AS 111, 195
\bibitem[1996]{heber96}
Heber, U., Dreizler, S., Hagen, H.-J., 1996, A\&A 311, L17
\bibitem[1998]{homeier}
Homeier, D., Koester, D., Hagen, H.-J., et al., 1998, A\&A, 338, 563
\bibitem[1975]{illarionov}
Illarionov, A.F., Sunyaev, R.A., 1975, A\&A 39, 185
\bibitem[1997]{kilkenny97}
Kilkenny, D., O'Donoghue, D., Koen, C., Stobie, R.S., Chen, A., MNRAS 287, 867
\bibitem[1969]{jaidee69}
Jaidee, S., Lyng\aa , G., 1969, Arkiv f. Astronomi 5, 345
\bibitem[2000]{lamontagne00}
Lamontagne, R., Demers, S., Wesemael, F., Fontaine, G., Irwin, M.J., 
2000, ApJ 119, 241
\bibitem[1998]{lemke98}
Lemke, M., Heber, U., Dreizler, S., Napiwotzki, R., Engels, D.: 1997, 
New results from the stellar component of the Hamburg Schmidt 
survey: A sample of sdO stars. In: The third conference on Faint 
Blue Stars, A.G.D. Philip, J. Liebert, R.A. Saffer (eds.), 
L. Davis Press, Schenectady, N.Y., 375
\bibitem[1999]{mccook}
McCook, G.P., Sion, E.M., 1999, ApJS 121, 1
\bibitem[1999]{napi}
Napiwotzki, R., Green, P.J., Saffer, R.A., 1999, ApJ 517, 399
\bibitem[1986]{paquette}
Paquette, C., Pelletier, C., Fontaine, G., Michaud, G., 1986, ApJS 61, 197
\bibitem[1992]{press}
Press, W.H., Teukolsky, S.A., Vetterling, W.T., Flannery, B.P., 1992, 
Numerical Recipes in FORTRAN, 2nd Edition, Cambridge University 
Press, Cambridge
\bibitem[1994]{reimersa}
Reimers, D., Jordan, S., K\"ohler, T., Wisotzki, L., 1994, A\&A 285, 995
\bibitem[1996]{reimersb}
Reimers, D., Jordan, S., Koester, D., Bade, N., K\"ohler, T., 
Wisotzki, L., 1996, A\&A 311, 572
\bibitem[1998]{reimersc}
Reimers, D., Jordan, S., Beckmann, V., Christlieb, N., 
Wisotzki, L., 1998, A\&A 337, L13 
\bibitem[1997]{reimers97}
Reimers, D., Wisotzki, L., 1997, The Messenger 88, 14
\bibitem[1999]{sefako99}
Sefako, R.R., Glass, I.S., Kilkenny, D., et al., 1999, MNRAS 309, 1043
\bibitem[1987]{shipman87}
Shipman, H.L., Liebert, J., Green, R.F., 1987, ApJ 315, 239
\bibitem[1997]{stobie97}
Stobie, R.S., Kilkenny, D., O'Donoghue, D., et al., 1997, MNRAS 287, 848
\bibitem[1982]{wesemael82}
Wesemael, F., Truran, J.W., 1982, ApJ 260, 807
\bibitem[1996]{wisotzki}
Wisotzki, L., K\"ohler, T., Groote, D., Reimers, D., 1996, A\&AS 115, 
227 
\bibitem[2000]{wisotzki00}
Wisotzki, L., Christlieb, N., Bade, N., Beckmann, V., K\"ohler, T., 
Vanelle, C., Reimers, D., 2000, in press 

\end{thebibliography}
\end{document}